\documentclass[final,number,sort&compress,5p,times,twocolumn]{elsarticle}

\journal{Nucl.\ Instr.\ and Meth.\ A}

\begin{document}

\begin{frontmatter}

\title{A Large Area LaBr$_3$/NaI Phoswich for Hard X-ray Astronomy}

\author{Hong Li}

\author{Jianfeng Ji\corref{cor}}
\ead{jijianf@tsinghua.edu.cn}
\cortext[cor]{Corresponding author}

\author{Hua Feng}

\author{Zhi Zhang}

\author{Dong Han}

\address{Department of Engineering Physics and Center for Astrophysics, Tsinghua University, Beijing 100084, China}
\address{Key Laboratory of Particle \& Radiation Imaging (Tsinghua University), Ministry of Education, China}

\begin{abstract}

In terms of energy resolution, temporal response to burst events, and thermal stability, lanthanum bromide doped with Ce is a much better choice than the traditional NaI(Tl) scintillator for hard X-ray astronomy. We present the test results of a phoswich detector with a diameter of 101.6~mm consisting of 6~mm thick LaBr$_3$:Ce and 40~mm thick NaI(Tl), which is the largest one of this type reported so far. The measured energy resolution is 10.6\% at 60 keV, varying inversely proportional to the square root of the energy, and the energy nonlinearity is found to be less than 1\%, as good as those of smaller phoswiches. The coupled scintillators and phototube also show excellent uniformity across the detecting surface, with a deviation of 0.7\% on the pulse amplitude produced by 60~keV gamma-rays. Thanks to the large ratio of light decay times of NaI(Tl) and LaBr$_3$:Ce, 250~ns vs.\ 16~ns, pulse shape discrimination is much easier for this combination than for NaI(Tl)/CsI(Na). As the light decay time of LaBr$_3$:Ce is about 15 times faster than that of NaI(Tl), this phoswich is more suitable for detection of bright, transient sources such as gamma-ray bursts and soft gamma-ray repeaters. The internal activity of lanthanum produces a count rate of about 6~counts~s$^{-1}$ at 37.5~keV in the detector. This peak could be used for in-flight spectral calibration and gain correction.

\end{abstract}

\begin{keyword}
LaBr$_3$ \sep NaI \sep phoswich \sep X-ray \sep gamma-ray \sep astronomy \sep burst
\end{keyword}

\end{frontmatter}

\section{Introduction}

The phoswich detector has the advantages of high photopeak efficiency and outstanding background rejection capability over a single scintillator. The most traditional configuration of a phoswich is a combination of NaI(Tl) and CsI(Na) crystals, with NaI being the sensitive volume and CsI for anti-coincidence.  This has been widely used in hard X-ray and soft gamma-ray astronomy over the past decades, especially where large detecting area, fast timing, and/or low background are required, e.g., the Phoswich Detection System (PDS) \citep{fro97} onboard BeppoSAX and the High Energy X-ray Timing Experiment \citep{rot98} onboard the Rossi X-ray Timing Explorer. Phoswiches like NaI/CsI are also selected for use in future space missions such as the Hard X-ray Modulation Telescope \citep{lit07} and the Space-based multi-band astronomical Variable Object Monitor \citep{don10}. Other types of phoswich are also in use, such as the GSO(Ce)/BGO hard X-ray detector on Suzaku \citep{kam96,tak07}. Therefore, investigation of new, advanced phoswiches would be interesting in X-ray astronomy.

Multiple crystals with distinct light decay times coupling to a single photomultiplier tube (PMT) provides a simple and reliable readout configuration, in contrast to high-Z semiconductor arrays of the same effective area, and is thus suitable for observations where the need for large area is prior to that for position sensitivity. However, such a large format of detector in front of a single readout has a potential disadvantage that pulses pileup is more likely to occur given the same input flux density, which will limit the maximum count rate that one can measure with the detector \citep{mee09}. The dead time of scintillators is limited by its light decay time. The best way to overcome this issue without complicating the system is to find new crystals with shorter decay times.

Recently, lanthanum bromide crystals doped with cerium (LaBr$_3$:Ce) has been extensively studied and used due to its excellent physical properties. The energy resolution of LaBr$_3$:Ce is found to be around 3\% at 662~keV \citep{van01}, which, versus 7\% for NaI(Tl), is a great improvement and is even comparable to the resolution of semiconductor (e.g.\ CdZnTe) detectors typical of 1--2\%. LaBr$_3$:Ce has a decay time as fast as 16~ns, more than 10 times shorter than that of NaI(Tl). Therefore, LaBr$_3$:Ce should be a better choice than NaI(Tl) for applications in hard X-ray astronomy as mentioned above. Also, its excellent thermal stability \citep{mos06} and high tolerance to radiation damage \citep{nor07,owe07} make it suitable and robust for space applications. Balloon flights for X-ray astronomy with a single LaBr$_3$:Ce crystal have been carried out \citep{man11}.

\citeauthor{maz10} \citep{maz10} reported characteristics of a phoswich assembly with LaBr$_3$:Ce and NaI(Tl), with a diameter of 3~inch (= 76.2~mm). \citeauthor{man11} \citep{man11} mentioned that a similar phoswich had been designed and under study. In this paper, we report on the test of a 4~inch (= 101.6~mm) diameter LaBr$_3$:Ce/NaI(Tl) phoswich (\S~\ref{sec:det}), about its temporal and spectral characteristics, uniformity (\S~\ref{sec:char}), and internal radioactivity (\S~\ref{sec:int}). A brief discussion and summary of the results is presented in \S~\ref{sec:diss}.

\section{Detector}
\label{sec:det}

\begin{figure}[t]
\centering
\includegraphics[width=\columnwidth]{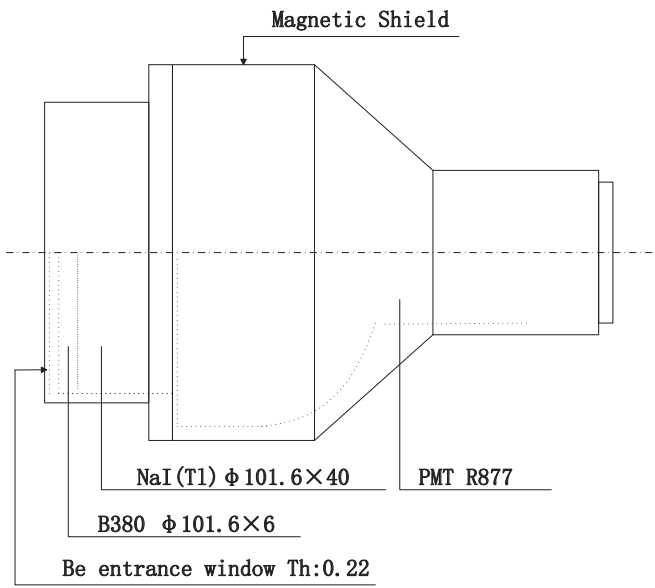}
\caption{Schematic drawing of the LaBr$_3$:Ce/NaI(Tl) phoswich.
\label{fig:det}}
\end{figure}

The phoswich consists of cylindrical LaBr$_3$:Ce and NaI(Tl) with a diameter of 101.6~mm. The LaBr$_3$:Ce is 6 mm thick and the NaI(Tl) is 40~mm thick. Such a design allows a full-energy detection efficiency of 20\% up to 300~keV. The entrance is shielded by a 0.22~mm thick Be window above LaBr$_3$:Ce, which allows at least 90\% throughput for photons above 6~keV. A PMT (Hamamatsu R877) that has a bialkali window of 127~mm in diameter and magnetic shield is mounted against the NaI(Tl). The whole detector was manufactured by the Saint-Gobain group corporate using their BrilLanCe 380 and NaI(Tl) crystals. Figure~\ref{fig:det} shows the schematic drawing of the phoswich detector.

The PMT was operated at a negative high voltage of $-$860~V. Signals output from the PMT anode were fed into a preamplifier with a discharge time scale of 100~ns. A pulse shape discrimination based on the front-rear-edge technique \citep{wu95} was applied to distinguish from which crystal the signal arose. All the measurements were made at room temperature.

\section{Temporal and Spectral Responses}
\label{sec:char}

\begin{figure}
\centering
\includegraphics[width=\columnwidth]{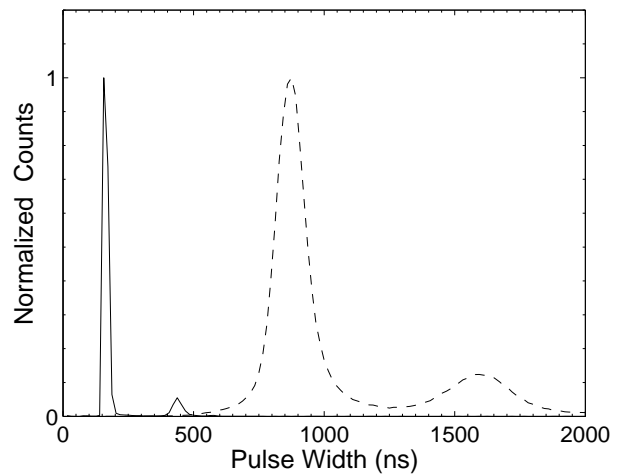}
\caption{
Pulse width spectrum measured with the LaBr$_3$:Ce/NaI(Tl) and NaI(Tl)/CsI(Na) phoswiches and with an $^{241}$Am source. The solid line is for LaBr$_3$:Ce/NaI(Tl), where the left peak corresponds to events from the LaBr3:Ce crystal and the right peak is for NaI(Tl). The dashed line is for NaI(Tl)/CsI(Na), where the left peak is for NaI(Tl) and the right is for Cs(Tl). Each curve is normalized to unity at the peak.
\label{fig:time}}
\end{figure}

The decay time of LaBr$_3$:Ce is 16~ns versus 250~ns for NaI(Tl). Distinct pulse widths are expected for energy deposits in the two crystals. Figure~\ref{fig:time} shows the pulse width spectrum from signals obtained by irradiating the phoswich with an $^{241}$Am source. The spectrum is characterized by two well separated peaks, one centered at 157~ns and the other at 439~ns, corresponding to energy deposits in LaBr$_3$:Ce and NaI(Tl), respectively. This is consistent with previous results obtained from a smaller LaBr$_3$:Ce/NaI(Tl) phoswich \citep{maz10}. Energy deposits in LaBr$_3$:Ce can then be identified by screening pulse widths. For comparison, we also presented a similar measurement with a NaI(Tl)/CsI(Na) phoswich of similar size \citep{jij07}, see the dashed curve in Figure~\ref{fig:time}.

\begin{figure*}
\centering
\includegraphics[width=\textwidth]{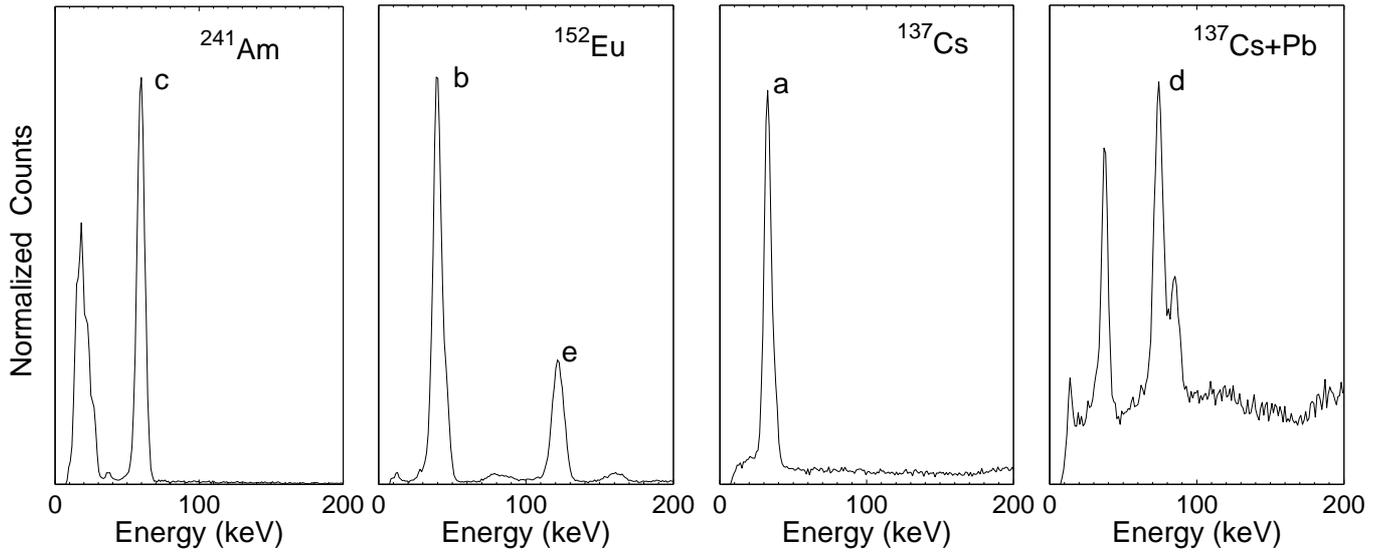}
\caption{
Energy spectra of $^{241}$Am, $^{152}$Eu, $^{137}$Cs, and lead irradiated by $^{137}$Cs, measured with the LaBr$_3$:Ce/NaI(Tl) phoswich. The five photopeaks used for spectral calibrations are at energies of (a) 32.1~keV (Ba K$\alpha$ from $^{137}$Cs), (b) 39.9~keV (Sm K$\alpha$ from $^{152}$Eu), (c) 59.5~keV~($^{241}$Am $\gamma$), (d) 74.2~keV (Pb K$\alpha$),and (e) 121.8~keV~($^{152}$Eu $\gamma$). All spectra are in linear scale and normalized to have the same maximum.
\label{fig:spec}}
\end{figure*}

To investigate linearity of such a large phoswich, a variety of radioactive sources were used. Figure~\ref{fig:spec} shows energy spectra of $^{241}$Am, $^{152}$Eu, $^{137}$Cs, and lead irradiated by $^{137}$Cs, measured with the detector. Only events with a pulse width in the LaBr$_3$:Ce peak are selected. We fitted each photopeak with a Gaussian function, plus a polynomial component accounting for the continuum if necessary. The measured energy resolution in full width at half maximum (FWHM) are 5.1, 6.0, 6.3, 7.4, and 9.1~keV, respectively, at 32.1~keV (Ba K$\alpha$ from $^{137}$Cs), 39.9~keV (Sm K$\alpha$ from $^{152}$Eu), 59.5~keV~($^{241}$Am), 74.2~keV (Pb K$\alpha$), and 121.8~keV~($^{152}$Eu). This is nearly a factor of 2 better than the resolution of NaI(Tl)/CsI(Na) phoswich of a comparable size \citep{kur75,fro97}.

\begin{figure}[h!]
\centering
\includegraphics[width=0.8\columnwidth]{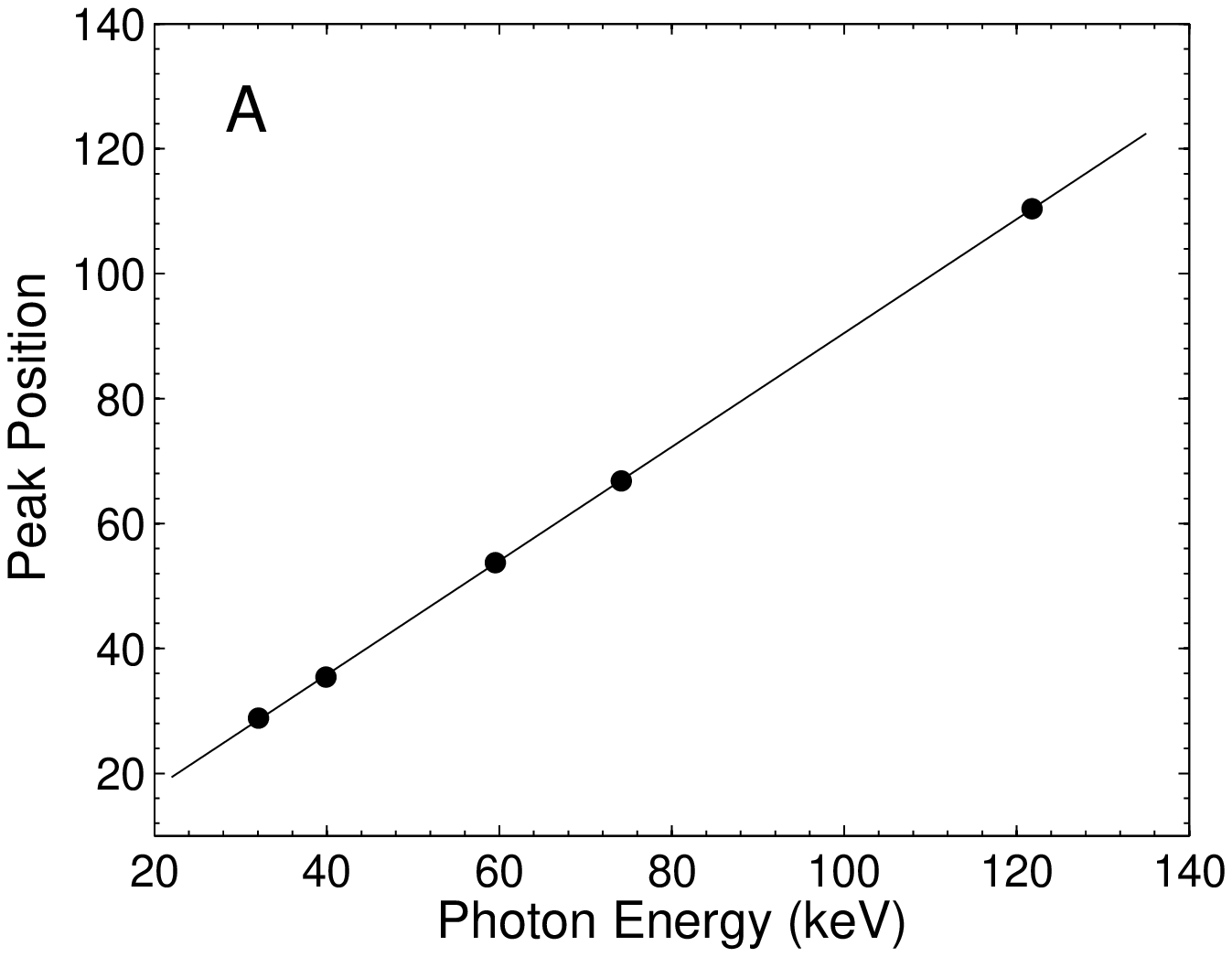}\\
\includegraphics[width=0.8\columnwidth]{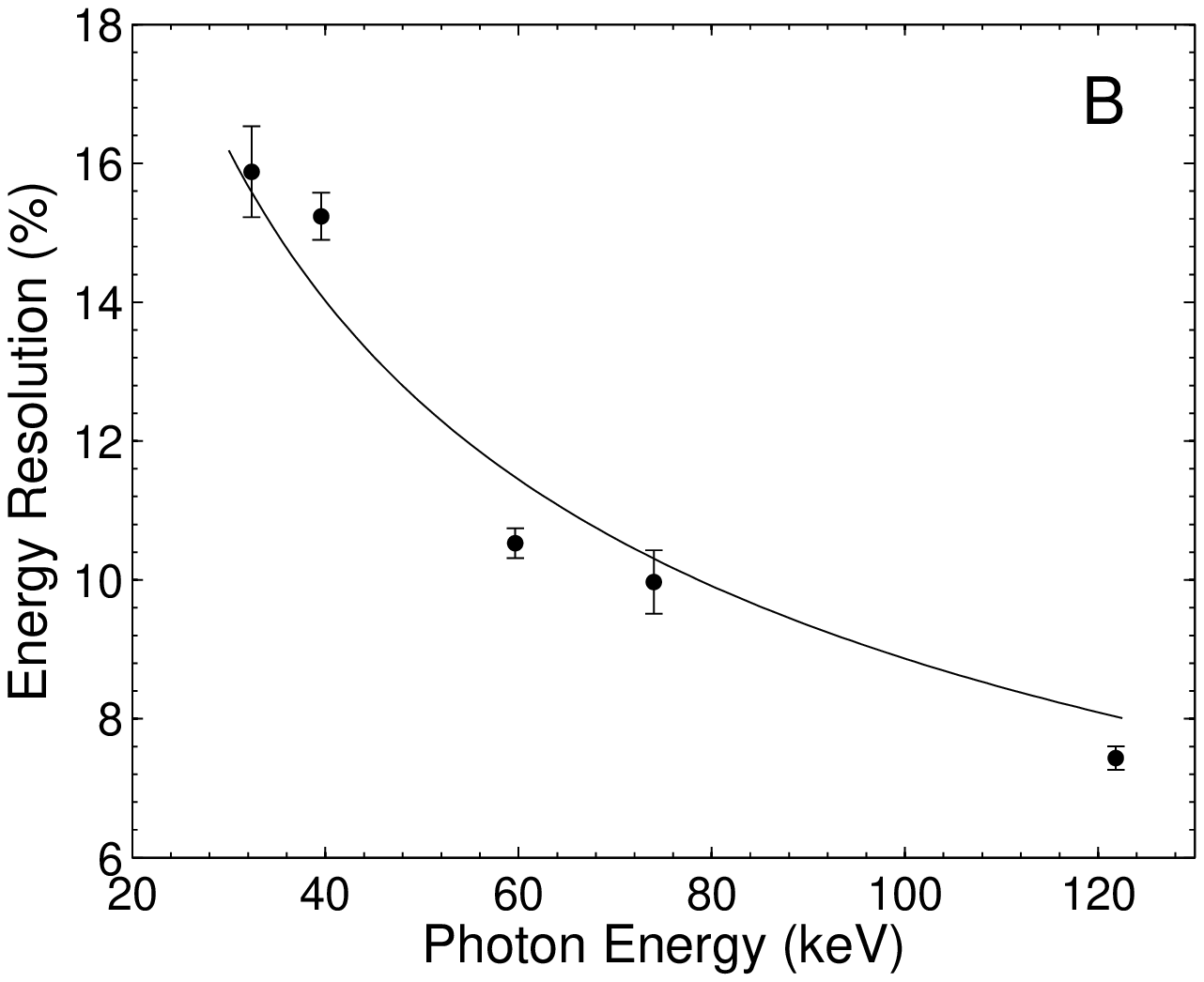}
\caption{
Peak channel position (A) and resolution in FHWM/$E$ (B) versus energy of the LaBr$_3$:Ce/NaI(Tl) phoswich, measured with radioactive sources shown in Figure~\ref{fig:spec}.  The error on the peak position is smaller than the marker. The energy nonlinearity is measured to be less than 1\%. The best-fit energy resolution curve has a form of FWHM$/E = 0.88 E^{-0.5}$.
\label{fig:linres}}
\end{figure}

\begin{figure}[t]
\centering
\includegraphics[width=\columnwidth]{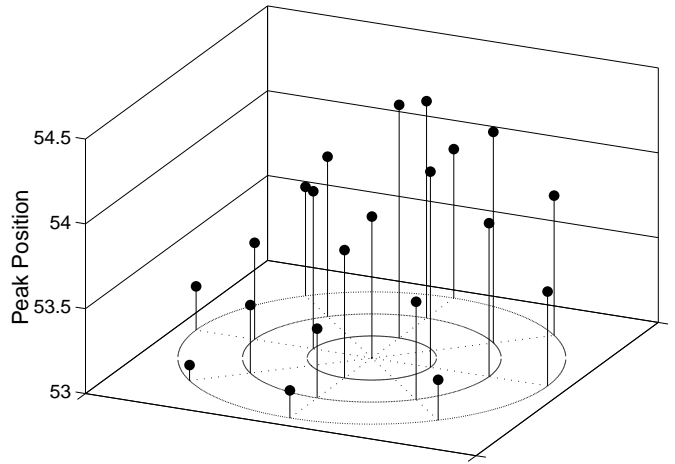}
\caption{
The channel centroids of the 59.54~keV gamma-ray photopeak varies as a function of the position over the detector surface. The peak position has a fractional standard deviation of 0.7\% over the surface.
\label{fig:uni}}
\end{figure}

\begin{figure}[t]
\centering
\includegraphics[width=\columnwidth]{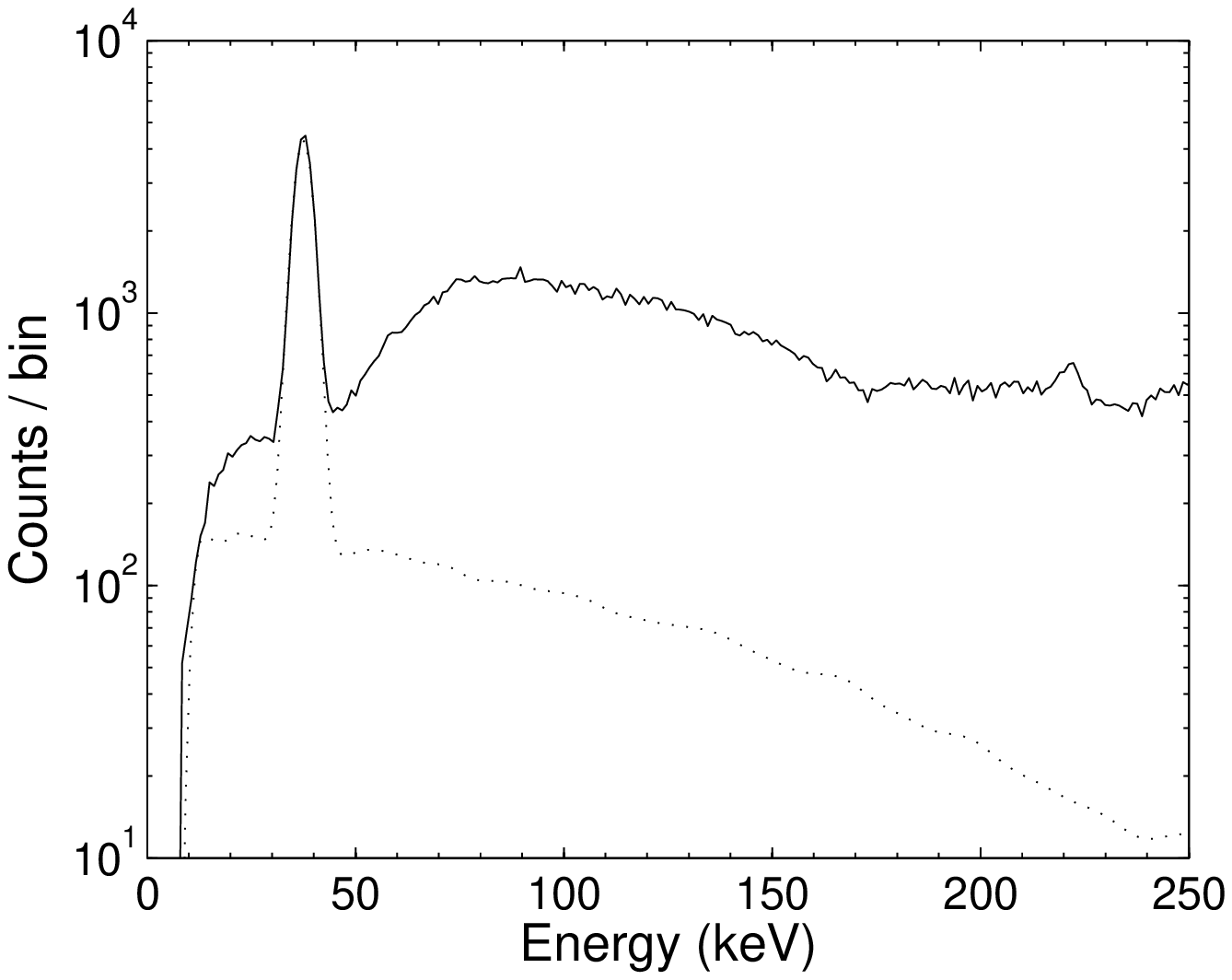}
\caption{
Laboratory background spectrum measured with the LaBr$_3$:Ce/NaI(Tl) phoswich. The peak at 37.45~keV, consistent with the Ba K shell binding energy, is caused by internal emission of a series of barium X-rays or auger electrons. The solid curve is the measured spectrum with an accumulation time of 3600~s in units of counts per energy bin ($\approx 1.1$~keV). The dashed curve indicates the contribution from the internal background caused by decays of $^{138}$La, estimated using simulations with GEANT4.
\label{fig:back}}
\end{figure}

With five photopeaks over the energy range of 30--130~keV, the nonlinearity of the phoswich detector was measured to be less than 1\%, see Figure~\ref{fig:linres}a. The fractional energy resolution versus energy is shown in Figure~\ref{fig:linres}b; a $E^{-0.5}$ function can adequately fit the curve, with $\chi^2 = 3.8$ and 4 degrees of freedom. The best-fit fractional energy resolution is FWHM$/E = 0.88(E/{\rm keV})^{-0.5}$. This is consistent with the results obtained from a relatively smaller single LaBr$_3$:Ce scintillator \citep{man11}.

The uniformity of spectral responses across the detecting surface is an important property of large area detectors. For scintillators, differences on light yielding, transmission, reflection, and collection at various spots may contribute to the nonuniformity. A poor uniformity would greatly degrade the spectral resolution of large area detectors. Here we only focus on the overall uniformity of whole system. We selected 21 positions evenly across the detecting surface, and exposed them to an $^{241}$Am source with a single hole collimator of 5~mm in diameter. The channel centroids of the photopeak from 59.54~keV gamma-rays at each position are shown in Figure~\ref{fig:uni}. The peak centroids has a fractional standard deviation of 0.7\% over the surface.

\section{Internal radioactivity}
\label{sec:int}

$^{138}$La is a natural radioisotope of lanthanum with a half-life of $1.02 \times 10^{11}$~year. It decays to $^{138}$Ba by electron capture with a chance of 66.4\%, or $^{138}$Ce by electron emission with a chance of 33.6\%. The former branch produces a 1436~keV gamma-ray followed by characteristic X-rays of barium, and the latter branch results in emission of a 789~keV gamma-ray and an electron with an endpoint energy of 255~keV. Due to the presence of $^{138}$La, the total internal activity of LaBr$_3$:Ce is estimated to be 1.6~Bq~cm$^{-3}$, which can be observed especially in large format crystals. In the energy band of interest, the barium fluorescences can be detected as a strong peak and contribute to the background. \citeauthor{maz10} \citep{maz10} concluded that, based on their phoswich geometry, 20\% of the barium peak can be rejected by anti-coincidence of the simultaneous 1436~keV gamma-ray. This peak in the background spectrum, in return, can be picked out using a similar coincidence technique and used for in-orbit calibration and gain correction.

Figure~\ref{fig:back} shows a spectrum of the laboratory background measured with our phoswich detector with an accumulation time of 3600~s. A barium peak is significantly detected at a count rate of 6.1~counts~s$^{-1}$. The central energy of the peak is measured to be $37.45 \pm 0.15$~keV. This is significantly larger than the Ba K$\alpha$ energy of 32.1~keV, but consistent with the binding energy of the Ba K shell.  This peak in some literatures \citep{maz10} has been misidentified as a Ba K$\alpha$ line at 32~keV, but instead, it should also include energies of cascade transitions following the K$\alpha$ or auger emission. In order to estimate the contribution by the internal radioactivity, we ran simulations with GEANT4 and the interal background spectrum is shown as a dashed curve in Figure~\ref{fig:back}. In our energy range of interest, the internal background by $^{138}$La mainly contributes to the peak around 37~keV and the low energy part of the spectrum.

\section{Discussion and Summary}
\label{sec:diss}

This is so far the largest LaBr$_3$:Ce/NaI(Tl) phoswich reported in the literature. The detector presents an energy resolution of 10.6\% at 60 keV and varies following the $1/\sqrt{E}$ relation, indicating that the energy uncertainty is dominated by Fano fluctuations. The pulse width spectrum shows two distinct peaks respectively from energy deposits in LaBr$_3$:Ce and NaI(Tl). Compared with the NaI(Tl)/CsI(Na) phoswich, whose decay time ratio is $630/250 \approx 2.5$, the two peaks are more widely separated due to a larger time ratio of $250/16 \approx 17$. This makes the pulse shape discrimination much easier, and will lead to a lower probability of misidentification. The lowest detectable energy of the detector is estimated to be around 2--3~keV, limited by the electronics noise and also the Be window; a Ti K$\alpha$ line at 4.5~keV can be detected distinctly. The variation of the pulse amplitude is only about 0.7\% across the sensitive surface of the detector; this is much smaller than the energy fluctuation and its contribution to the spectral resolution is negligible in this energy band. The detector performance, based on our tests, is found to be as good as previously measured for smaller, single LaBr$_3$:Ce crystals, indicating that the growth and sealing techniques required for a large phoswich of this type is ready.

In principle, this phoswich can achieve a dead time $\sim$10 times smaller than the NaI(Tl)/CsI(Na) phoswich can do. The maximal allowed count rate is inversely proportional to the dead time. Thus, the LaBr3:Ce crystal can measure a count rate 10 times higher than NaI(Tl) can without pulses pileup. This would bring interesting applications in detecting burst events such as the gamma-ray bursts and soft gamma-ray repeaters. Despite the excellent performance of LaBr$_3$:Ce/NaI(Tl), its high internal activity seems to be an issue for applications where low background is needed. This does not affect the detection of burst events, which occur at a short duration with high flux and thus the instrumental background can be ignored. For long gamma-ray bursts that last up to $\sim$100 seconds, the internal background may become significant and should be taken into account. In the 200--300~keV band, the internal background level of our detector estimated from simulations is about 0.3~counts~s$^{-1}$, or 0.0037~counts~s$^{-1}$~cm$^{-2}$. Above 300~keV, the internal background spectrum is almost flat in units of counts per keV. For comparison, the orbital background of BeppoSAX/PDS\footnote{See Table 6 in section 4.6.3.3 of the BeppoSAX handbook available at ftp://ftp.asdc.asi.it/sax/doc/handbook/} has a count rate of 0.014~counts~s$^{-1}$~cm$^{-2}$ in 200--300~keV. We note that the background of BeppoSAX/PDS is lower than many other space instruments due to its low inclination orbit. Thus, the internal background by $^{138}$La is a minor or negligible component compared with the total orbital background in the hard band above 200~keV, and will not affect the detection of the high energy tail in relatively long astronomical phenomena. Interesting applications may include measuring the peak energy and its temporal evolution of long gamma-ray bursts \citep{fro00}, or the search of hard X-ray emission in the afterglow \citep{tka00,mai05}. While in the soft band below 200 keV, the estimated internal background is about 1.6--2 times of the orbital background of PDS, and will reduce the sensitivity by a factor of $\sim$1.7. On the other hand, the electron capture of $^{138}$La with a pair of 37~keV and 1436~keV events allows coincident measurement of the 37~keV peak, which can be used for in-orbit calibration and gain correction. For that purpose, a proper design of the readout is needed, to precisely measure both the fast, low energy event in LaBr$_3$:Ce, and a slow, high energy event in NaI(Tl). This is under investigation in our lab and will be reported in the future.

\section*{Acknowledgements}

We thank Elena Pian for helpful discussions about GRB detection and the anonymous referee for useful comments. This work is partially supported by the National Natural Science Foundation of China under grant No.\ 10903004 and 10978001, and the 973 Program of China under grant 2009CB824800.


\begin{thebibliography}{00}

\bibitem[Frontera et al.(1997)]{fro97}
F.\ Frontera, E.\ Costa, D.\ dal Fiume, M.\ Feroci, L.\ Nicastro, M.\ Orlandini, E.\ Palazzi, G.\ Zavattini, Astron.\ Astrophys.\ Suppl.\ 122 (1997) 357.

\bibitem[Rothschild et al.(1998)]{rot98}
R.E.\ Rothschild, P.R.\ Blanco, D.E.\ Gruber, W.A.\ Heindl, D.R.\ MacDonald, D.C.\ Marsden, M.R.\ Pelling, L.R.\ Wayne, R.L.\ Hink, Astrophys.\ J.\ 496 (1998) 538.

\bibitem[Li(2007)]{lit07}
T.P.\ Li, 2007, Nucl.\ Phys.\ B Proc.\ Suppl.\ 166 (2007) 131.

\bibitem[Dong et al.(2010)]{don10}
Y.\ Dong, B.\ Wu, Y.\ Li, Y.\ Zhang, S.\ Zhang, Sci.\ Ch.\ Phys.\ 53 (2010) 40.

\bibitem[Kamae et al.(1996)]{kam96}
T.\ Kamae et al., SPIE Proc., 2806 (1996) 314.

\bibitem[Takahashi et al.(2007)]{tak07}
T.\ Takahashi et al., Publ.\ Astron.\ Soc.\ Jpn. 59 (2007) 35.

\bibitem[Meegan et al.(2009)]{mee09}
C.\ Meegan et al., Astrophys.\ J., 702 (2009) 791.

\bibitem[van Loef et al.(2001)]{van01}
E.V.D.\ van Loef, P.\ Dorenbos, C.W.E.\ van Eijk, Appl.\ Phys.\ Lett.\ 79 (2001) 1573.

\bibitem[Moszy{\'n}ski et al.(2006)]{mos06}
M.\ Moszy{\'n}ski et al.\, Nucl.\ Instr.\ and Meth.\ A 568 (2006) 739

\bibitem[Normand et al.(2007)]{nor07}
S.\ Normand, A.\ Iltis, F.\ Bernard, T.\ Domenech, P.\ Delacour, P., Nucl.\ Instr.\ and Meth.\ A 572 (2007) 754.

\bibitem[Owens et al.(2007)]{owe07}
A.\ Owens, A.J.J.\ Bos, S.\ Brandenburg, E.J.\ Buis, C.\ Dathy, P.\ Dorenbos, C.W.E.\ van Eijk, S.\ Kraft, R.W.\ Ostendorf, V.\ Ouspenski, F.\ Quarati, Nucl.\ Instr.\ and Meth.\ A 572 (2007) 785.

\bibitem[Manchanda(2011)]{man11}
R.K.\ Manchanda, Adv. Space Res.\ 47 (2011) 30.

\bibitem[Mazumdar et al.(2010)]{maz10}
I.\ Mazumdar, G.A.\ Kumar, D.A.\ Gothe, R.K.\ Manchanda, Nucl.\ Instr.\ and Meth.\ A 623 (2010) 995.

\bibitem[Wu et al.(1995)]{wu95}
M.\ Wu, L.\ Cheng, Y.\ Li, J.\ Wang, H.\ Wang, B.\ Wu, C.\ Zhang, Nucl.\ Electron.\ Detect.\ Technol.\ 15 (1995) 13.

\bibitem[Ji et al.(2007)]{jij07}
J.\ Ji, Z.\ Zhang, C.\ Liu, Nucl.\ Electron.\ Detect.\ Technol.\ 27 (2007) 458.

\bibitem[Kurfess \& Johnson(1975)]{kur75}
J.D.\ Kurfess, W.N.\ Johnson, IEEE Trans.\ Nucl.\ Sci. 22 (1975) 626.

\bibitem[Frontera et al.(2000)]{fro00}
F.\ Frontera et al., Astrophys.\ J.\ Suppl.\ Ser.\ 127 (2000) 59.

\bibitem[Tkachenko et al.(2000)]{tka00}	
A.Y.\ Tkachenko, O.V.\ Terekhov, R.A.\ Sunyaev, R.A.\ Burenin, C.\ Barat, J.-P.\ Dezalay, G.\ Vedrenne, Astron.\ Astrophys. 358 (2000) L41.

\bibitem[Maiorano et al.(2005)]{mai05}
E.\ Maiorano et al., Astron.\ Astrophys.\ 438 (2005) 821.

\end{thebibliography}
\end{document}